\documentclass[twocolumn]{autart}    

\usepackage{graphicx}          

\usepackage{amsmath}
\usepackage{amssymb}
\usepackage{algorithmicx}
\usepackage{algpseudocode}

\usepackage{IEEEtrantools}

\newtheorem{theorem}{Theorem}
\newtheorem{lemma}{Lemma}
\newtheorem{proposition}{Proposition}
\newtheorem{corollary}{Corollary}

\begin{document}

\begin{frontmatter}

\title{Maximizing the Smallest Eigenvalue of a Symmetric Matrix: A Submodular Optimization Approach} 

\thanks[footnoteinfo]{A preliminary version of this paper was presented at the American Control Conference (ACC) 2017~\cite{clark2017submodular}. Corresponding author A. Clark. Tel. 1-508-831-6962.}

\author[WPI]{Andrew Clark}\ead{aclark@wpi.edu}, 
\author[WPI]{Qiqiang Hou}\ead{qhou@wpi.edu}, 
\author[UW]{Linda Bushnell}\ead{lb2@uw.edu}, and 
\author[UW]{Radha Poovendran}\ead{rp3@uw.edu}

\address[WPI]{Worcester Polytechnic Institute, Worcester, MA, USA 01609}
\address[UW]{University of Washington, Seattle, WA, USA 98195-2500}



\begin{abstract}                          
This paper studies the problem of selecting a submatrix of a positive definite matrix in order to achieve a desired bound on the smallest eigenvalue of the submatrix. Maximizing this smallest eigenvalue has applications to selecting input nodes in order to guarantee consensus of networks with negative edges as well as maximizing the convergence rate of distributed systems. We develop a submodular optimization approach to maximizing the smallest eigenvalue by first proving that positivity of the eigenvalues of a submatrix can be characterized using the probability distribution of the quadratic form induced by the submatrix. We then exploit that connection to prove that positive-definiteness of a submatrix can be expressed as a constraint on a submodular function. We prove that our approach results in polynomial-time algorithms with provable bounds on the size of the submatrix. We also present generalizations to non-symmetric matrices, alternative sufficient conditions for the smallest eigenvalue to exceed a desired bound that are valid for Laplacian matrices, and a numerical evaluation.
\end{abstract}

\end{frontmatter}

\section{Introduction}
\label{sec:intro}


An increasingly widespread approach to controlling networked systems is to select a set of nodes to perform actuation (e.g., selecting generators to participate in power system control, or designating agents as leaders in multi-agent systems), while relying on network effects to steer the remaining nodes to a desired state~\cite{liu2011controllability,tanner2004controllability}. Mathematically, this approach is often modeled as creating an induced submatrix, in which rows and columns corresponding to the leaders are removed~\cite{barooah2007estimation}. The dynamics of the remaining network nodes are then specified by the induced submatrix. A prominent example of this class of systems is the grounded Laplacian matrix, which is created in consensus networks when the states of a subset of leader nodes are set identically to zero \cite{pirani2014spectral}. 

The performance of such systems is known to be heavily influenced by the spectrum of the induced submatrix \cite{pirani2016smallest}. Of particular importance is the smallest eigenvalue of the induced sub-matrix. In \cite{rahmani2009controllability}, it was shown that the rate of convergence of a consensus network is determined by the magnitude of the smallest eigenvalue of the grounded Laplacian matrix. The sign of the smallest eigenvalue determines whether the system is stable. Networks with antagonistic interactions, such as biological regulatory networks with repressive connections or social networks in which users disagree, may be unstable~\cite{chen2016characterizing,zelazo2017robustness}. Ensuring consensus in such systems is equivalent to selecting a submatrix in which all eigenvalues are positive.


   
   
   The role of the smallest eigenvalue of the grounded matrix implies that an analytical approach to selecting submatrices in which all eigenvalues are bounded below by a desired value would lead to improved stability and faster convergence of networked systems \cite{pirani2016smallest,rahmani2009controllability,zelazo2017robustness}. However, so far to the best of our knowledge there are no computational techniques for maximizing the smallest eigenvalue. The main difficulty is that, unlike metrics such as the inverse trace \cite{clark2014supermodular} and convergence error \cite{clark2014minimizing}, the smallest eigenvalue of the grounded Laplacian is not known to possess any structure such as submodularity that enables development of efficient optimization algorithms with formal guarantees. 
   Hence, while an efficient input selection algorithm with provable guarantees would improve the stability, robustness, and convergence rate of networked systems, at present no such algorithms that maximize the smallest eigenvalue are available.
   
   In this paper, we present a submodular optimization approach to input selection in order to maximize the smallest eigenvalue of an induced submatrix such as the grounded Laplacian. Specifically, we investigate the problem of selecting a minimum-size input set in order to guarantee that the smallest eigenvalue is above a desired threshold. Our approach is as follows. We first prove that the eigenvalue condition holds if and only if an induced quadratic form is positive with probability one. Second, we show that this condition can be mapped to a constraint on a submodular function, equal to the probability that the quadratic form is zero when the input is a Gaussian random vector. Finally, we prove that this probability can be computed in polynomial time. 
   
   We analyze the optimality guarantees of our proposed approach and prove that it the number of selected input nodes is within a logarithmic bound of the minimum-size input set.  We show that the submodular optimization approach is applicable to problems including ensuring consensus of signed networks and maximizing convergence rate, and also explore generalizations to non-symmetric matrices (e.g., arising from directed graphs). We propose alternative sufficient conditions that are applicable to Laplacian matrices. Our sufficient conditions consist of bounds on the inverse trace and log determinant of the submatrix, and are shown to be submodular via spectral submodularity techniques. Our approach is validated through numerical study. 
   
   The paper is organized as follows. Section \ref{sec:related} reviews the related work. Section \ref{sec:preliminaries} gives relevant background. Section \ref{sec:formulation} presents the problem formulation and two motivating applications. Section \ref{sec:submodular} presents our proposed submodular framework. Section \ref{sec:extension} discusses extensions to non-symmetric matrices and alternative sufficient conditions. Section \ref{sec:simulation} contains numerical results. Section \ref{sec:conclusion} concludes the paper.
\section{Related Work}
\label{sec:related}
The importance of the smallest eigenvalue of grounded Laplacian graphs was identified in \cite{rahmani2009controllability}, where it was shown that the magnitude of the smallest eigenvalue determines the rate of convergence to consensus. The eigenvalues of the grounded Laplacian were further studied in \cite{pirani2014spectral,pirani2016smallest}. While these works analyzed the impact of the smallest eigenvalue and developed bounds on the smallest eigenvalue for different classes of graph, the problem of selecting nodes based on this criterion remains open.

Consensus in networks with both positive and negative edge weights, in which the negative weights represent antagonistic interactions between nodes, has been studied in \cite{alemzadeh2017controllability,zelazo2014definiteness}. Necessary and sufficient conditions for consensus in such networks  without inputs based on effective resistance were proposed in \cite{chen2016characterizing,zelazo2014definiteness}. To the best of our knowledge, the only work that considers input selection in order to ensure consensus in networks with negative edges is the preliminary conference version of this paper \cite{clark2017submodular}. Compared to \cite{clark2017submodular}, this paper presents tighter necessary and sufficient conditions for consensus. The related problem of controllability of signed networks was proposed in \cite{alemzadeh2017controllability}, but makes fundamentally different assumptions, namely that the input nodes can follow any arbitrary state trajectory.

The performance of networked systems with input nodes, often denoted as leaders, has been studied extensively~\cite{jadbabaie2003coordination,liu2011controllability,tanner2004controllability}. In particular, prior works have proposed techniques for selecting input nodes to optimize metrics including robustness to noise~\cite{clark2014supermodular}, convergence rate~\cite{clark2014minimizing}, and controllability~\cite{summers2016submodularity}, with submodular optimization as one approach. At present, however, there are no polynomial-time algorithms with provable guarantees for selecting input nodes in order to optimize the minimum eigenvalue of networked systems.

\section{Notation and Preliminaries}
\label{sec:preliminaries}
In what follows, we give needed background on symmetric matrices, probability, and submodularity, and define notations that will be used throughout the paper.

Let $I_{n}$ denote the $n \times n$ identity matrix. We omit the subscript $n$ when the dimensionality of the matrix is clear for compactness of notation. A matrix $A$ is symmetric if $A = A^{T}$, where $A^{T}$ denotes the transpose of $A$. Any symmetric matrix can be written in the form $A = U\Lambda U^{T}$, where $U$ is a unitary matrix (i.e., $UU^{T} = I$) and $\Lambda$ is a real diagonal matrix.  A symmetric matrix $A$ is positive definite if all eigenvalues are positive, or equivalently, if $v^{T}Av > 0$ for all vectors $v$. The notation $A \succ 0$ denotes positive definiteness of $A$, while $A \succ B$ if $(A-B)$ is positive definite. For any matrix $A$, the set of eigenvalues of $A$ is denoted as $\lambda_{1}(A),\ldots,\lambda_{n}(A)$, where it is assumed that $\lambda_{1}(A) \geq \lambda_{2}(A) \geq  \cdots \geq \lambda_{n}(A)$. We also use the notation $\lambda_{min}(A)$ to denote the minimum eigenvalue of $A$. Finally, we let $D(S)$ denote a diagonal matrix with $(D(S))_{ii} = 1$ if $i \in S$ and all other entries $0$.

For an $n \times n$ matrix $A$, let $S \subseteq \{1,\ldots,n\}$ denote a set of indices. We let $A(S)$ denote the submatrix formed by the rows and columns indexed in $S$. Some interpretations of $S$ and $A(S)$ are discussed in Section \ref{sec:formulation}. The following theorem describes the relationship between the eigenvalues of a matrix $A$ and the eigenvalues of a submatrix.

\begin{theorem}[Cauchy Interlacing Theorem \cite{horn2012matrix}]
\label{theorem:CIT}
Let $A$ be an $n \times n$ symmetric matrix and let $A^{\prime} = A(\{1,\ldots,n\} \setminus \{i\})$ for some $i \in \{1,\ldots,n\}$. Then 
\begin{multline*}
\lambda_{1}(A) \geq \lambda_{1}(A^{\prime}) \geq \lambda_{2}(A) \geq \cdots \geq \lambda_{n-1}(A) \\
\geq \lambda_{n-1}(A^{\prime}) \geq \lambda_{n}(A).
\end{multline*}
\end{theorem}

As a corollary to Theorem \ref{theorem:CIT}, we have that if $S \subseteq T \subseteq \{1,\ldots,n\}$, then $\lambda_{min}(A(S)) \leq \lambda_{min}(A(T))$, or in other words, the minimum eigenvalue $\lambda_{min}(S)$ is monotone increasing in the set $S$.

We now define notations and basic properties for certain random variables. Throughout the paper, we let $f_{Z}(z)$ and $F_{Z}(z)$ denote the probability density and distribution functions of random variable $Z$ evaluated at $z \in \mathbb{R}$, respectively. We let $\mathbf{E}(\cdot)$ denote expectation, and let $Pr(\cdot)$ denote the probability of an event occurring. 

Recall that for a Gaussian random vector $\mathbf{z}$, with mean vector $\boldsymbol{\mu}$ and covariance matrix $\Sigma$, the random variable $M\mathbf{z}$ for any matrix $M$ is Gaussian with mean $M\boldsymbol{\mu}$ and covariance $M\Sigma M^{T}$. If $X_{1},\ldots,X_{r}$ are independent Gaussian random variables with zero mean and unit variance, then the random variable $Z = X_{1}^{2} + \cdots + X_{r}^{2}$ is a chi-squared random variable with $r$ degrees of freedom, with probability density function $$f_{Z}(z) = \frac{z^{\frac{r}{2}-1}e^{-\frac{z}{2}}}{2^{\frac{r}{2}}\Gamma\left(\frac{r}{2}\right)}$$ for $z > 0$ and $0$ otherwise, where $\Gamma$ denotes the gamma function. The mean of $Z$ is $r$, while the variance of $Z$ is $2r$.

Finally, we give brief background on submodular functions. Let $V$ denote a finite set. A function $f: 2^{V} \rightarrow \mathbb{R}$ that takes as input a subset of $V$ and gives as output a real number is \emph{submodular} if, for any sets $S,T \subseteq V$, $$f(S) + f(T) \geq f(S \cap T) + f(S \cup T).$$ Equivalently, $f$ is submodular if and only if, for any $S \subseteq T \subseteq V$ and any $v \notin T$, $$f(S \cup \{v\}) - f(S) \geq f(T \cup \{v\}) - f(T).$$ A function $f$ is supermodular if $-f$ is submodular, while a function is modular if it is both submodular and supermodular.  Any positive weighted linear combination of submodular functions is submodular. Finally, the following lemma gives a further construction of submodular functions.
\begin{lemma}[\cite{fujishige2005submodular}]
\label{lemma:submod_min}
Suppose that $g(S)$ is nondecreasing and submodular as a function of $S$. Then for any real number $\zeta$, the function $$\overline{g}(S) = \min{\{f(S), \alpha\}}$$ is nondecreasing and submodular as a function of $S$.
\end{lemma}

\section{Problem Formulation and Motivation}
\label{sec:formulation}
This section presents the problem formulation, as well as motivating applications to maximizing the convergence rate of a leader-follower network and ensuring consensus in a network with negative edge weights. 

Let $A$ denote a symmetric $n \times n$ matrix, and let $V = \{1,\ldots,n\}$. The problem studied in this paper is formulated as

\begin{equation}
\label{eq:formulation}
\begin{array}{ll}
\mbox{minimize} & |S| \\
\mbox{s.t.} & \lambda_{min}(A(V \setminus S)) \geq \beta
\end{array}
\end{equation}
where $\beta \in \mathbb{R}$. In words, Eq. (\ref{eq:formulation}) seeks to remove the minimum-size set $S$ of rows and columns of a matrix $A$ in order to ensure that the eigenvalues of the sub-matrix $A(V \setminus S)$ are above a bound $\beta$. We will prove that (\ref{eq:formulation}) is equivalent to a submodular optimization problem. First, however, we will motivate (\ref{eq:formulation}) by discussing its connection to consensus problems.

Consider a network of $n$ nodes, indexed in the set $V = \{1,\ldots,n\}$. An edge $(i,j)$ between nodes $i$ and $j$ exists if node $i$ influences the dynamics of node $j$ and vice versa. Edges are assumed to be undirected, and the set of edges is denoted $E$. The neighbor set of node $i$ is defined as $N_{i} = \{j : (i,j) \in E\}$, and consists of the set of nodes that influence the dynamics of $i$. There is a nonnegative weight $W_{ij}$ for each edge $(i,j) \in E$, with $W_{ij} = W_{ji}$.

For such a network, we define the Laplacian matrix $L$ as the $n \times n$ symmetric matrix with entries 
\begin{displaymath}
L_{ij} = \left\{
\begin{array}{ll}
-W_{ij}, & (i,j) \in E \\
\sum_{j \in N(i)}{W_{ij}}, & i = j \\
0, & \mbox{else}
\end{array}
\right.
\end{displaymath}

Each node $i$ has a time-varying real-valued state $x_{i}(t)$. The input nodes maintain  constant state values, which are assumed to be zero without loss of generality. The non-input nodes have state dynamics 
\begin{equation}
\label{eq:consensus}
\dot{x}_{i}(t) = -\sum_{j \in N_{i}}{W_{ij}(x_{i}(t)-x_{j}(t))}.
\end{equation}
 In this application, the set $S$ is equivalent to the set of input nodes. To see this, let $\mathbf{x}_{S}(t)$ denote the state vector of the non-input nodes. The dynamics of the non-input nodes can then be written as $$\dot{\mathbf{x}}_{S}(t) = - L(V \setminus S)\mathbf{x}_{S}(t).$$ It was shown in \cite{jadbabaie2003coordination} that  $\mathbf{x}_{S}(t)$ will converge to zero, provided that each non-input node is path-connected to at least one input node. The following proposition, which is analogous to a result first demonstrated in \cite{rahmani2009controllability}, describes the convergence rate of the dynamics (\ref{eq:consensus}).

\begin{proposition}
\label{prop:consensus_converges}
The state vector $\mathbf{x}_{S}(t)$ satisfies $$||\mathbf{x}_{S}(t)||_{2} \leq e^{-\lambda_{min}(L(V \setminus S))t}||\mathbf{x}_{S}(0)||_{2}.$$
\end{proposition}

The proof follows from straightforward Lyapunov analysis using the function $V(\mathbf{x}_{S}) = \frac{1}{2}\mathbf{x}_{S}^{T}\mathbf{x}_{S}$. Hence, maximizing the minimum eigenvalue will minimize the convergence rate to consensus, and choosing the set $S$ in Eq. (\ref{eq:formulation}) is equivalent to selecting a set of input nodes that satisfy a given bound on the convergence rate.   Practical applications include maximizing the speed of influence propagation in a social network~\cite{ghaderi2013opinion}, as well as improving the performance of formation control algorithms, which often use consensus as an inner loop \cite{ren2008distributed}.

The consensus dynamics (\ref{eq:consensus}) can also be considered in networks where $W_{ij} < 0$ for some edges $(i,j)$. Such negative weights represent antagonistic interactions between nodes, for example, negative social interactions~\cite{li2013influence} or repressive regulation in biological networks. In such networks, convergence to consensus is not guaranteed because the Laplacian matrix may not be positive definite, and the problem of selecting a subset of input nodes in order to ensure consensus in a signed network is equivalent to problem (\ref{eq:formulation}) with $\beta = 0$.


\section{Proposed Submodular Optimization Approach}
\label{sec:submodular}

This section presents our submodular optimization approach to maximize the smallest eigenvalue of a submatrix (Eq. (\ref{eq:formulation})). The proposed approach is valid for any symmetric matrix $A$, including matrices representing networks with negative edges. We first present an equivalent problem formulation to (\ref{eq:formulation}) and prove that it is submodular. We then propose algorithms that exploit the submodular structure and analyze their complexity and optimality bounds.

\subsection{Equivalent Formulation and Proof of Submodularity}
\label{subsec:submodular}
We first observe that we can assume that $\beta = 0$ in (\ref{eq:formulation}) without loss of generality, since we can construct a new matrix $\hat{A} = A - \beta I$ if needed and ensure that $\lambda_{min}(\hat{A}(S)) > 0$. Our equivalent formulation arises from the following preliminary lemma.

\begin{lemma}
\label{lemma:probability}
For any $n \times n$ symmetric matrix $A$ and subset $S \subseteq \{1,\ldots,n\}$, the following are equivalent:
\begin{enumerate}
\item[(i)] $A(S)$ is positive definite.
\item[(ii)] There exists $\alpha > 0$ such that $A + \alpha D(V\setminus S)$ is positive definite.
\item[(iii)] If $\mathbf{w}$ is an $n$-dimensional Gaussian random vector with mean $0$ and covariance matrix $I$, then $$\mathbf{E}\left(\min{\left\{\alpha\sum_{i \notin S}{w_{i}^{2}} + w^{T}Aw,0\right\}}\right) = 0.$$
\end{enumerate}
\end{lemma}

\emph{Proof:}
We first show that (i) and (ii) are equivalent. Since $A(S)$ is a submatrix of $A + \alpha D(S)$, we have that $A + \alpha D(S) \succ 0$ implies that $A(S) \succ 0$. Now, suppose that $A(S) \succ 0$, and suppose without loss of generality that $S = \{1,\ldots,k\}$ for some $k$. Then $A + \alpha D(V \setminus S) \succ 0$ is equivalent to
\begin{displaymath}
\left(
\begin{array}{cc}
A(S) & A(S,V\setminus S) \\
A(S,V \setminus S) & A(V \setminus S, V \setminus S) + \alpha I
\end{array}
\right) \succ 0.
\end{displaymath}
By the Schur complement theorem, $A + \alpha D(V \setminus S) \succ 0$ if and only if $A(S) \succ 0$, which is true by assumption, and $$A(V \setminus S, V \setminus S) + \alpha I \succ A(S, V \setminus S)A(S)^{-1}A(V \setminus S, S),$$ which can be satisfied by choosing $\alpha$ sufficiently large.

We now show that (ii) and (iii) are equivalent. First, if (ii) holds, then $$\min{\left\{\alpha\sum_{i \notin S}{w_{i}^{2}} + w^{T}Aw,0\right\}} = 0$$ for all $w$, and hence the expectation is zero. Conversely, suppose that (iii) holds, and yet $A + \alpha D(V \setminus S)$ is not positive definite. Let $\mathbf{x} \in \mathbb{R}^{n}$ satisfy $\mathbf{x}^{T}(A + \alpha D(V \setminus S))\mathbf{x} = -\delta$ for some $\delta > 0$. Since $\mathbf{x}^{T}(A + \alpha D(V \setminus S))\mathbf{x}$ is a continuous function of $\mathbf{x}$, there is a ball $B(\mathbf{x},\epsilon)$ centered on $\mathbf{x}$ with radius $\epsilon$ such that $\mathbf{y}^{T}(A + \alpha D(V \setminus S))\mathbf{y} < - \frac{\delta}{2}$ for all $\mathbf{y} \in B(\mathbf{x},\epsilon)$. Hence
\begin{IEEEeqnarray*}{rCl}
\IEEEeqnarraymulticol{3}{l}{
\mathbf{E}\left(\min{\left\{\alpha\sum_{i \notin S}{w_{i}^{2}} + w^{T}Aw,0\right\}}\right)} \\ 
&\leq& \int_{B(\mathbf{x},\epsilon)}{\min{\left\{\alpha\sum_{i \notin S}{y_{i}^{2}} + \mathbf{y}^{T}A\mathbf{y},0\right\}}f_{\mathbf{w}}(\mathbf{y}) \ d\mathbf{y}} \\
&<& 0,
\end{IEEEeqnarray*}
a contradiction. \qed

We define the function $Q(S)$ as $$Q(S) \triangleq \mathbf{E}\left(\min{\left\{\mathbf{w}^{T}(A + \alpha D(S))\mathbf{w}, 0\right\}}\right),$$ where $\mathbf{w}$ is an $N(\mathbf{0},I)$ random vector of dimension $n$. By Lemma \ref{lemma:probability}, Eq. (\ref{eq:formulation}) is equivalent to
\begin{equation}
\label{eq:equiv_form}
\begin{array}{ll}
\mbox{minimize} & |S| \\
\mbox{s.t.} & Q(S) = 0
\end{array}
\end{equation}

The following theorem establishes that (\ref{eq:equiv_form}) is a submodular optimization problem.

\begin{theorem}
\label{theorem:Q_supermodular}
The function $Q(S)$ is increasing and submodular as a function of $S$.
\end{theorem}

\emph{Proof:} The function $Q(S)$ is an integral that can be approximated as a limit of Riemann sums. Define $Q_{m}(S)$ as 
\begin{multline*}
Q_{m}(S) \\
= \sum_{i=1}^{m}{\min{\left\{(x^{mi})^{T}(A + \alpha D(S))x^{mi},0\right\}}f_{\mathbf{w}}(x^{mi})\delta_{m}},
\end{multline*}
 where the $\{x^{mi}:  i=1,\ldots,m, \ m = 1,2,\ldots\}$ and $\{\delta_{m}: m=1,2,\ldots\}$ are chosen so that $Q_{m}(S)$ converges to $Q(S)$ as $m \rightarrow \infty$.

For each $m$ and $i$, we have that $$(x^{mi})^{T}(A + \alpha D(S))x^{mi} = (x^{mi})^{T}Ax^{mi} + \alpha\sum_{j \in S}{(x_{j}^{mi})^{2}},$$ which is increasing and modular as a function of $S$. Hence by Lemma \ref{lemma:submod_min}, $$\min{\left\{(x^{mi})^{T}(A + \alpha D(S))x^{mi}, 0\right\}}$$ is an increasing submodular function of $S$.  $Q_{m}(S)$ is therefore a nonnegative weighted sum of increasing submodular functions, and hence is increasing and submodular. 

Finally, $Q(S)$ is the pointwise limit of a sequence of submodular functions $Q_{m}(S)$, and hence is submodular. \qed

Intuitively, formulation (\ref{eq:equiv_form}) can be interpreted as a covering constraint, namely, for every vector $x$ satisfying $x^{T}Ax < 0$, there must exist an $i \in S$ such that $x_{i} \neq 0$. Such covering problems are typically submodular. Since in this case the number of such vectors $x$ is uncountably infinite, we instead require  that a Gaussian random vector $\mathbf{w}$ satisfies $\mathbf{w}^{T}A\mathbf{w} > 0$ with probability $1$, which is equivalent by Lemma \ref{lemma:probability}. 

Theorem \ref{theorem:Q_supermodular} implies that a submodular optimization approach can be used to approximate (\ref{eq:equiv_form}) with provable optimality guarantees. These techniques are described in detail in the following section. 

\subsection{Algorithms and Analysis}
\label{subsec:algorithms}

This section presents algorithms that exploit the submodular structure identified in Theorem \ref{theorem:Q_supermodular} to approximate the solution to (\ref{eq:equiv_form}) with provable optimality bounds. We first discuss computation of $Q(S)$, and then present an algorithm for approximating (\ref{eq:equiv_form}).

The computation of $Q(S)$ is as follows. First, we write $A + \alpha D(S) = U\Lambda U^{T}$, where $U$ is a unitary matrix and $\Lambda$ is diagonal. Hence if $\mathbf{w}$ is an $N(0,I)$ Gaussian random variable, then $$\mathbf{w}^{T}(A + \alpha D(S))\mathbf{w} = \mathbf{z}^{T}\Lambda\mathbf{z},$$ where $\mathbf{z}$ is a Gaussian random variable with zero mean and covariance matrix $UU^{T} = I$, i.e., a vector of independent standard normal random variables. The random variable  $\mathbf{z}^{T}\Lambda\mathbf{z}$ is a linear combination of $\chi^{2}$ random variables. Letting $Z = \mathbf{z}^{T}\Lambda\mathbf{z}$, $Q(S)$ can be expressed as $$Q(S) = \int_{-\infty}^{0}{z f_{Z}(z) \ dz}.$$ Equivalently, if we define $\hat{Z} = -Z$, then $$Q(S) = -\int_{0}^{\infty}{\hat{z}f_{\hat{Z}}(\hat{z}) \ d\hat{z}} = -\int_{0}^{\infty}{Pr(\hat{Z} > \hat{z}) \ d\hat{z}}.$$ $Q(S)$ can be computed via numerical integration of $Pr(\hat{Z} > \hat{z})$. The following result gives an approach for computing the integrand.

\begin{proposition}[\cite{imhof1961computing}]
\label{prop:cdf-comp}
Let $W = a_{1}Y_{1} + \cdots + a_{m}Y_{m}$, where $a_{1},\ldots,a_{m}$ are scalars and $Y_{1},\ldots,Y_{m}$ are $\chi_{1}^{2}$-random variables with mean $1$. Then 
\begin{equation}
\label{eq:cdf}
Pr(W > w) = \frac{1}{2} + \frac{1}{\pi}\int_{0}^{\infty}{\frac{\sin{\theta(u)}}{u\rho(u)} \ du},
\end{equation}
 where 

\begin{displaymath}
\theta(u) = \frac{1}{2}\sum_{r=1}^{m}{(\tan^{-1}{(a_{r}u)})} - \frac{1}{2}xu, \quad 
\rho(u) = \prod_{r=1}^{m}{(1+a_{r}^{2}u^{2})^{\frac{1}{4}}}
\end{displaymath}

\end{proposition}

We define the following approximations for computing $Q(S)$. We let $g(z ; S) = Pr(\hat{Z} > z)$ and $\overline{g}(z; S,K)$ to be the truncation of (\ref{eq:cdf}) at $K > 0$, defined as $$\overline{g}(z ; S,K) = \frac{1}{2} + \frac{1}{\pi}\int_{0}^{K}{\frac{\sin{\theta(u)}}{u\rho(u)} \ du}.$$

\begin{proposition}
\label{prop:parameters}
For any $\epsilon > 0$, there exist parameters $R$, $K$,  and $N$ satisfying $R = O((-\log{\epsilon} + n)\lambda_{max}(A))$, $K = O(\frac{1}{\lambda_{min}(A)})$, and $N = O(R/\epsilon)$, such that $|Q(S) - \overline{Q}(S)| < \epsilon$.
\end{proposition}

\emph{Proof:} The expression $|Q(S)-\overline{Q}(S)$ can be bounded by  
\begin{IEEEeqnarray}{rCl}
\IEEEeqnarraymulticol{3}{l}{
\nonumber
|Q(S) - \overline{Q}(S)|} \\
\nonumber
 &=& \left|\int_{0}^{\infty}{g(z; S) \ dz} - \int_{0}^{R}{g(z ; S) \ dz} + \int_{0}^{R}{g(z ; S) \ dz} \right. \\
 \nonumber
 && \left. - \sum_{i=1}^{N}{g(z_{i},S)\delta} + \sum_{i=1}^{N}{g(z_{i} ; S)\delta} - \sum_{i=1}^{N}{\overline{g}(z_{i} ; S)\delta}\right| \\
\label{eq:complexity-main-1}
&\leq& \left|\int_{R}^{\infty}{g(z ; S) \ dz}\right| \\
\label{eq:complexity-main-2}
&& + \left|\int_{0}^{R}{g(z ; S) \ dz} - \sum_{i=1}^{N}{g(z_{i} ; S) \delta}\right| \\
\label{eq:complexity-main-3}
&& + \left|\sum_{i=1}^{N}{\delta(g(z_{i} ; S) - \overline{g}(z_{i} ; S))}\right|
\end{IEEEeqnarray}
We consider each term (\ref{eq:complexity-main-1})--(\ref{eq:complexity-main-3}) separately and show that the chosen parameters lead to an $O(\epsilon)$ error bound. For the first term, define $Z_{2} = |\lambda_{m+1}|w_{m+1} + \cdots + |\lambda_{n-|S|}|w_{n-|S|}$, where $\lambda_{m+1},\ldots,\lambda_{n-|S|}$ are the negative eigenvalues of $A(V \setminus S)$. We use a Chernoff bound to estimate  $g(z ; S)$ as 
\begin{displaymath}
Pr(\hat{Z} > z) \leq Pr(Z_{2} > z) \leq \min_{t \geq 0}{e^{-tz}\prod_{i=m+1}^{n-|S|}{\exp{(tw_{i})}}}. 
\end{displaymath}
The moment generating function of a $\chi_{1}^{2}$ random variable is equal to $$\Phi_{\chi_{1}^{2}}(t) = (1-2t)^{-1/2}, \quad t \in [0, 1/2].$$ Since each $W_{i}$ is equal to $\lambda_{i}Y_{i}$ where $Y_{i}$ is a $\chi_{1}^{2}$ random variable, the Chernoff bound can be simplified to 
\begin{eqnarray}
\nonumber
Pr(\hat{Z} > z) &=& \min_{t \in [0,\frac{1}{2\lambda_{max}}]}{e^{-tz}\prod_{i=1}^{n}{(1-2t\lambda_{i})^{-1/2}}} \\
\label{eq:chernoff}
&\leq& e^{-\frac{z}{4\lambda_{max}}}\prod_{i}{\left(1-\frac{\lambda_{i}}{2\lambda_{max}}\right)^{-1/2}},
\end{eqnarray}
where (\ref{eq:chernoff}) arises by setting $t = \frac{1}{4\lambda_{max}}$. 
We then have the bound 
\begin{multline*}
\int_{R}^{\infty}{g(z ; S) \ dz} \\
\leq \left(\prod_{i=1}^{n}{\left(1 - \frac{\lambda_{i}}{2\lambda_{max}}\right)^{-1/2}}\right)(4\lambda_{max}e^{-\frac{R}{4\lambda_{max}}}).
\end{multline*}
 Hence the choice of $R$ gives the desired bound on (\ref{eq:complexity-main-1}).

For (\ref{eq:complexity-main-2}), we have by \cite[Ch. 2.1]{davis1975methods} that $N = O(\frac{R}{\epsilon}) = O(\frac{n + \log{\epsilon}}{\epsilon^{2}})$ gives an $O(\epsilon)$ bound for the rectangular approximation to an integral over a closed interval. 

The error $|\overline{g}(z ; S) - g(z ; S)|$ is determined by the error in computing the integral (\ref{eq:cdf}). From \cite{imhof1961computing}, the approximation $\overline{g}(z ; S)$ has truncation error $T_{K}$ with $$T_{K}^{-1} \geq \frac{\pi m}{2} K^{m/2}\prod_{r=1}^{n}{|\lambda_{r}|^{1/2}},$$ and hence choosing $K = O(\frac{1}{\lambda_{min}})$ gives the desired error bound. \qed

Proposition \ref{prop:parameters} implies that $Q(S)$ can be approximated up to a bound $\epsilon$ through $O((-\log{\epsilon} + n)\lambda_{max}(A))$ evaluations of $\overline{g}(z ; S)$, each of which requires $O\left(\frac{R}{\epsilon\lambda_{min}}\right)$ computations. The greedy algorithm requires $O(n^{2})$ evaluations of $Q(S)$.

  


We next describe the algorithm for approximately solving (\ref{eq:formulation}), which uses the above described procedure for computing $Q(S)$ as a subroutine. The approximation algorithm is greedy, and is shown in pseudocode as Algorithm \ref{algo:greedy}. At each iteration, the algorithm selects the node $v$ that maximizes $Q(S \cup \{v\})$, terminating when $Q(S) = 0$.

   \begin{center}
\begin{algorithm}[!htp]
	\caption{Algorithm for selecting a set of input nodes to ensure that $Q(S) = 0$.}
	\label{algo:greedy}
	\begin{algorithmic}[1]
		\Procedure{Max\_Eigenvalue}{$A$}
            \State \textbf{Input}: Symmetric matrix $A$
            \State \textbf{Output}: Set of indices $S$
            \State Compute $\alpha$ as in Lemma \ref{lemma:probability}
            \State $S \leftarrow \emptyset$
             \While{$Q(S) < 0$} 
            \State $v^{\ast} \leftarrow \arg\min{\{Q(S \cup \{v\}): v \notin S\}}$
            \State $S \leftarrow S \cup \{v^{\ast}\}$
            \EndWhile
            \State \Return{$S$}
		\EndProcedure
	 \end{algorithmic}
\end{algorithm}
\end{center}

The optimality bound provided by Algorithm \ref{algo:greedy}, and the overall complexity, are described by the following proposition.

\begin{proposition}
Let $\hat{S}$ be the solution returned by Algorithm \ref{algo:greedy}, and let $S^{\ast}$ be the optimal solution to (\ref{eq:formulation}). Then 
\begin{equation}
\label{eq:optimality-bound}
\frac{|\hat{S}|}{|S^{\ast}|} \leq 1 + \log{\frac{Q(\emptyset)}{Q(S_{T-1})}}.
\end{equation}
\end{proposition}

\emph{Proof:} In \cite{wolsey1982analysis}, it was shown that the greedy algorithm returns a set $S$ satisfying $$\frac{|S|}{|S^{\ast}|} \leq 1 + \log{\left\{\frac{f(V)-f(\emptyset)}{f(V)-f(S_{T-1})}\right\}},$$ when $f(S)$ is submodular and  $S_{T-1}$ is the set at the second-to-last iteration of the algorithm. The proof then follows from submodularity of $Q(S)$ and the fact that $Q(\hat{S}) = 0$. \qed 

We observe that the bound in (\ref{eq:optimality-bound}) depends on the matrix $A$. Developing parameter-dependent bounds is a direction for future work.


\section{Extensions and Other Conditions}
\label{sec:extension}
This section describes extensions of our approach to non-symmetric matrices, and also gives two other sufficient conditions for bounding the eigenvalues. The first condition is valid for Laplacian matrices and is based on the trace of the inverse spectrum. The second condition is valid for any symmetric condition and is based on the log of the determinant. We show that both conditions are equivalent to submodular constraints.

\subsection{Extension to Non-Symmetric Matrices}
\label{subsec:non-symmetric}
We first explore a generalized problem in which the goal is to remove a subset of rows and columns of an arbitrary (not necessarily symmetric) matrix $A$ in order to ensure that $A(S)$ has only positive eigenvalues. By Lyapunov's Theorem, a matrix has eigenvalues with positive real parts if and only if there exists a positive definite matrix $P$ such that $A^{T}P + PA > 0$. A sufficient condition is that, for a diagonal matrix $D$ with positive entries, $A^{T}D + DA > 0$. The following lemma leads to our approach.

\begin{lemma}
\label{lemma:diagonal}
Let $D$ be a diagonal matrix, and let $B = A^{T}D + DA$. Then $B(S) = (A(S))^{T}D(S) + D(S)A(S)$.
\end{lemma}

\emph{Proof:} We have that $$B_{ij} = \sum_{l=1}^{n}{(A^{T})_{il}D_{lj}} + \sum_{l=1}^{n}{D_{il}A_{lj}} = A_{ji}D_{jj} + D_{ii}A_{ij}.$$ Hence the $(i,j)$ entry of $B(S)$ is determined entirely by $A(S)$ and $D(S)$, and so $B(S) = (A(S))^{T}D(S) + D(S)A(S)$ as desired. \qed

Our approach can therefore be extended to non-symmetric matrices by choosing a diagonal matrix $D$, constructing the matrix $B = A^{T}D + DA$, and then following Algorithm \ref{algo:greedy} to select a subset of rows and columns of $S$ that guarantee positive-definiteness. This corresponds to a submatrix of $A$ that has eigenvalues with positive real parts. We observe that while this condition is sufficient, it is not necessary since it is based on a specific choice of $D$, and the matrix $D$ is restricted to be diagonal instead of positive definite.

\subsection{Inverse Trace Conditions}
\label{subsec:alternative}
This section derives an alternative, spectral approach to selecting input nodes for the specific problem of ensuring consensus in a network with negative edges. For a Laplacian matrix $L$, we define $L = L_{+} - L_{-}$, where $L_{+}$ is the Laplacian induced by positive edges and $L_{-}$ is the Laplacian induced by negative edges. Letting $\zeta = |\lambda_{min}(L_{-})|$, we have the following initial result.

\begin{proposition}
\label{prop:effective-resistance-prelim}
If $\mathbf{trace}(L_{+}(S)^{-1}) \leq \frac{1}{\zeta}$, then the matrix $L(S)$ is positive definite.
\end{proposition}

\emph{Proof:} The condition $L(S) \prec 0$ is equivalent to $L_{+}(S) \prec L_{-}(S)$. Since $\lambda_{min}(L_{-}) \leq \lambda_{min}(L_{-}(S))$, a sufficient condition is $\lambda_{min}(L(S)) > \zeta$, or equivalently, $$\frac{1}{\lambda_{min}(L(S))} < \frac{1}{\zeta}.$$ 

Now, since $L_{+}(S)$ is symmetric and positive definite, $$\frac{1}{\lambda_{min}(L_{+}(S))} = \lambda_{max}((L_{+}(S))^{-1}).$$ Since $L_{+}(S)$ is positive definite, $L_{+}(S)^{-1}$ is positive definite, and hence 
\begin{eqnarray*}
\lambda_{max}((L_{+}(S))^{-1}) &\leq& \sum_{i=1}^{n}{\lambda_{i}((L_{+}(S))^{-1})} \\
&=& \mathbf{trace}((L_{+}(S))^{-1}),
\end{eqnarray*}
establishing the sufficient condition. \qed

A submodular approach to ensuring convergence based on the inverse trace is established by the following theorem.

\begin{theorem}[\cite{friedland2013submodular}]
\label{theorem:spectral}
Suppose that $A$ is a symmetric positive definite $M$-matrix, i.e., a symmetric positive definite matrix whose off-diagonal entries are negative. For any sets $S$ and $T$, 
\begin{multline*}
\mathbf{trace}(A(S)^{-1}) + \mathbf{trace}(A(T)^{-1}) \\ \leq \mathbf{trace}(A(S \cup T)^{-1}) + \mathbf{trace}(A(S \cap T)^{-1}).
\end{multline*}
\end{theorem}

Define $F(S) = \mathbf{trace}((L_{+}(S))^{-1})$. 

\begin{corollary}
\label{corollary:r-eff}
The function $F(S)$ is supermodular as a function of $S$.
\end{corollary}

\emph{Proof:} By construction, $L_{+}$ is a symmetric, positive definite matrix with negative off-diagonal entries, and hence is an M-matrix. Supermodularity then follows from Theorem \ref{theorem:spectral}. \qed

Proposition \ref{prop:effective-resistance-prelim} and Corollary \ref{corollary:r-eff} imply that selecting a set of rows and columns $S$ according to $$\min{\{|S| : F(S) \leq 1/|\lambda_{min}(L_{-})|\}}$$ is sufficient to ensure positive-definiteness of $L_{+}$, and hence convergence to consensus with negative edges. This problem can be solved approximately via an approach analogous to Algorithm \ref{algo:greedy}. Furthermore, this approach can be generalized to ensure convergence as a desired rate $\alpha$, as discussed in Section \ref{sec:formulation}, by considering the matrix $\tilde{L} = L -\alpha I$. Finally, the inverse trace can also be interpreted as the effective resistance with the grounded Laplacian \cite{clark2014supermodular}, thus establishing a connection between the submodular approach and the effective resistance-based characterizations of signed consensus found in \cite{chen2016characterizing,zelazo2014definiteness}.

\subsection{Determinant Conditions}
\label{subsec:determinant}

An additional sufficient condition can be established by using the following lemma, which relates the eigenvalues of a matrix to its determinant.

\begin{lemma}[\cite{merikoski1997bounds}]
\label{lemma:determinant}
Let $A$ be a symmetric positive definite matrix. Then the minimum eigenvalue of $A$ satisfies $$\lambda_{min}(A) \geq \left(\frac{n-1}{\mathbf{trace}(A)}\right)^{n-1}\det{(A)}.$$
\end{lemma}

Hence, a sufficient condition for positive definiteness is \begin{multline*}
 \left(\frac{n-1}{\mathbf{trace}(L + \alpha D(S) + \zeta I)}\right)^{n-1}\det{(L + \alpha D(S) + \zeta I)} \\
 \geq \zeta,
 \end{multline*}
which is equivalent to
\begin{multline}
\label{eq:determinant-condition}
\log{\det{(L + \alpha D(S) + \zeta I)}} \\
- (n-1) \log{(\mathbf{trace}(L) + \alpha |S| + \zeta n)} \\
> \log{\zeta} - (n-1) \log{(n-1)}.
\end{multline}
The following lemma explores the submodularity of this sufficient condition.
\begin{lemma}
\label{lemma:determinant-submodular}
The functions 
\begin{eqnarray*}
f_{1}(S) &=& \log{(\mathbf{trace}(L) + \alpha |S| + \zeta n)} \\
f_{2}(S) &=& \log{\det{(L + \alpha D(S) + \zeta I)}}
\end{eqnarray*}
are submodular.
\end{lemma}

\emph{Proof:} The function $\mathbf{trace}(L) + \alpha |S| + \zeta n$ is modular as a function of $S$. The function $f_{1}(S)$ is a composition of a concave and modular function, and is therefore submodular. Submodular of $f_{2}(S)$ follows from \cite{cortesi2014submodularity}. \qed  

Eq. (\ref{eq:determinant-condition}) is therefore equivalent to establishing a constraint on a difference of submodular functions. Efficient algorithms for approximating such problems have been studied in \cite{iyer2012algorithms}. Moreover, the function $f_{1}(S)$ has additional structure, namely, it depends only on the cardinality of $S$. Based on this structure, one approach is to find the smallest value of $k$ such that 
\begin{IEEEeqnarray*}{rCl}
\IEEEeqnarraymulticol{3}{l}{
\log{\det{(L + \alpha D(S) + \zeta I)}}}  \\
&>& (n-1)\log{(\mathbf{trace}(L) + \alpha k + \zeta n} \\
&& + \log{\zeta} - (n-1)\log{(n-1)}
\end{IEEEeqnarray*}
holds for some set $S$ with $|S| \leq k$. Such a set $S$ can be approximately obtained using a greedy algorithm analogous to Algorithm \ref{algo:greedy}. 
\section{Numerical Study}
\label{sec:simulation}
Our proposed submodular approach to eigenvalue maximization was evaluated numerically using Matlab. The simulation study considered the problem of ensuring consensus in a network with negative edges, introduced in Section \ref{sec:formulation}. The Laplacian matrices were generated as follows. 

We simulated a geometric random network in which an edge exists between two nodes if they are within a given distance of each other. The number of nodes varied from $20$ to $40$, with node positions set uniformly at random over an area with width chosen to ensure an average node degree of $4$. The range of each node was set to $300$. Each edge was chosen to have weight $1$ with probability $0.8$ and weight $-1$ with probability $0.2$. We investigated the effect of the number of nodes and the fraction of negative edges on the number of input nodes required for consensus, which is equivalent to the number of rows and columns that must be removed to ensure positive-definiteness. 

The submodular approach was compared with two heuristics. In the first heuristic, the rows and columns with the largest diagonal entries (corresponding to network nodes with maximum degree) were removed until positive-definiteness was achieved. In the second heuristic, rows and columns were removed randomly until positive-definiteness was achieved, or until no rows and columns remained.

\begin{figure*}
\centering
$\begin{array}{ccc}
\includegraphics[width=2in]{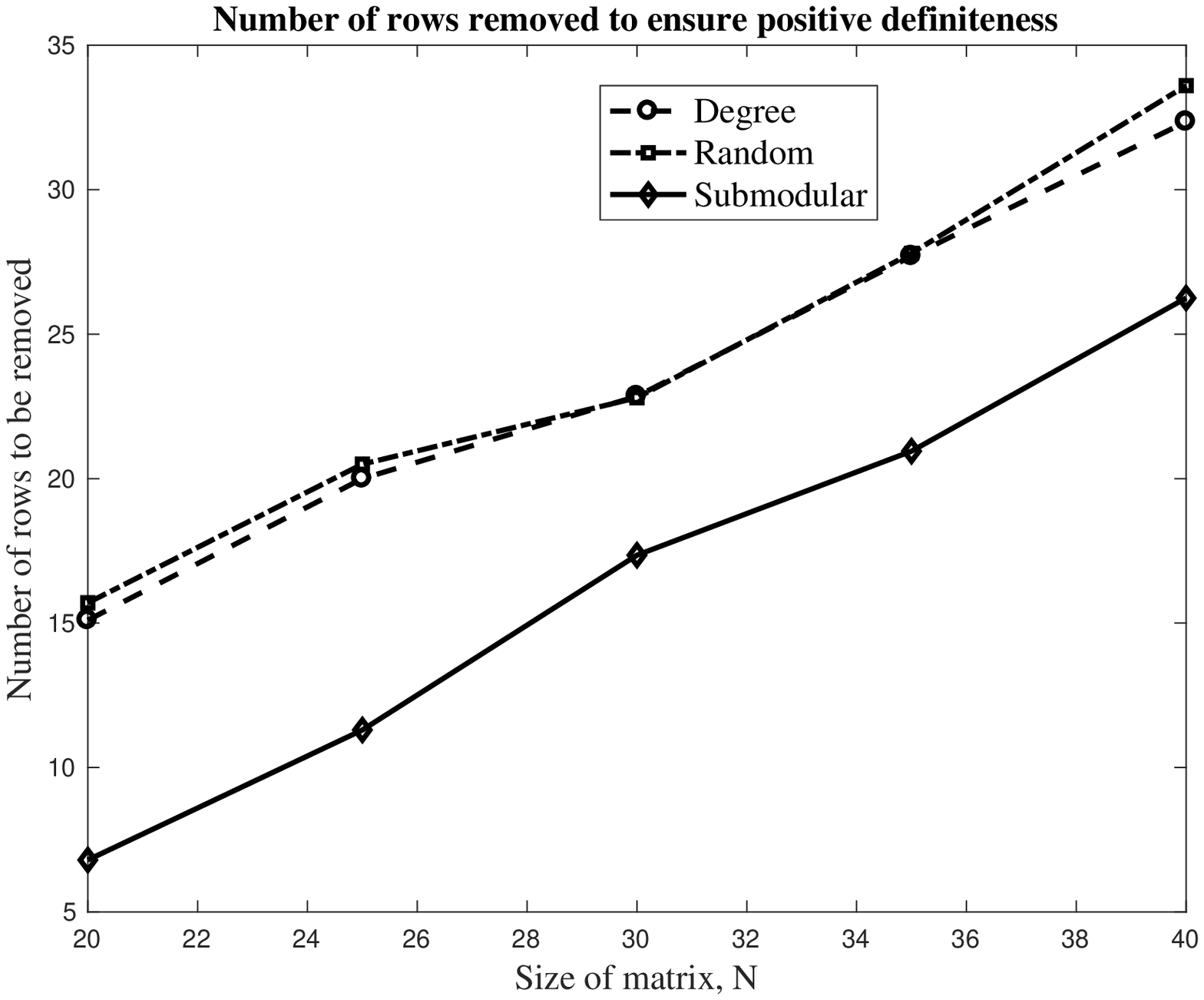} &
\includegraphics[width=2in]{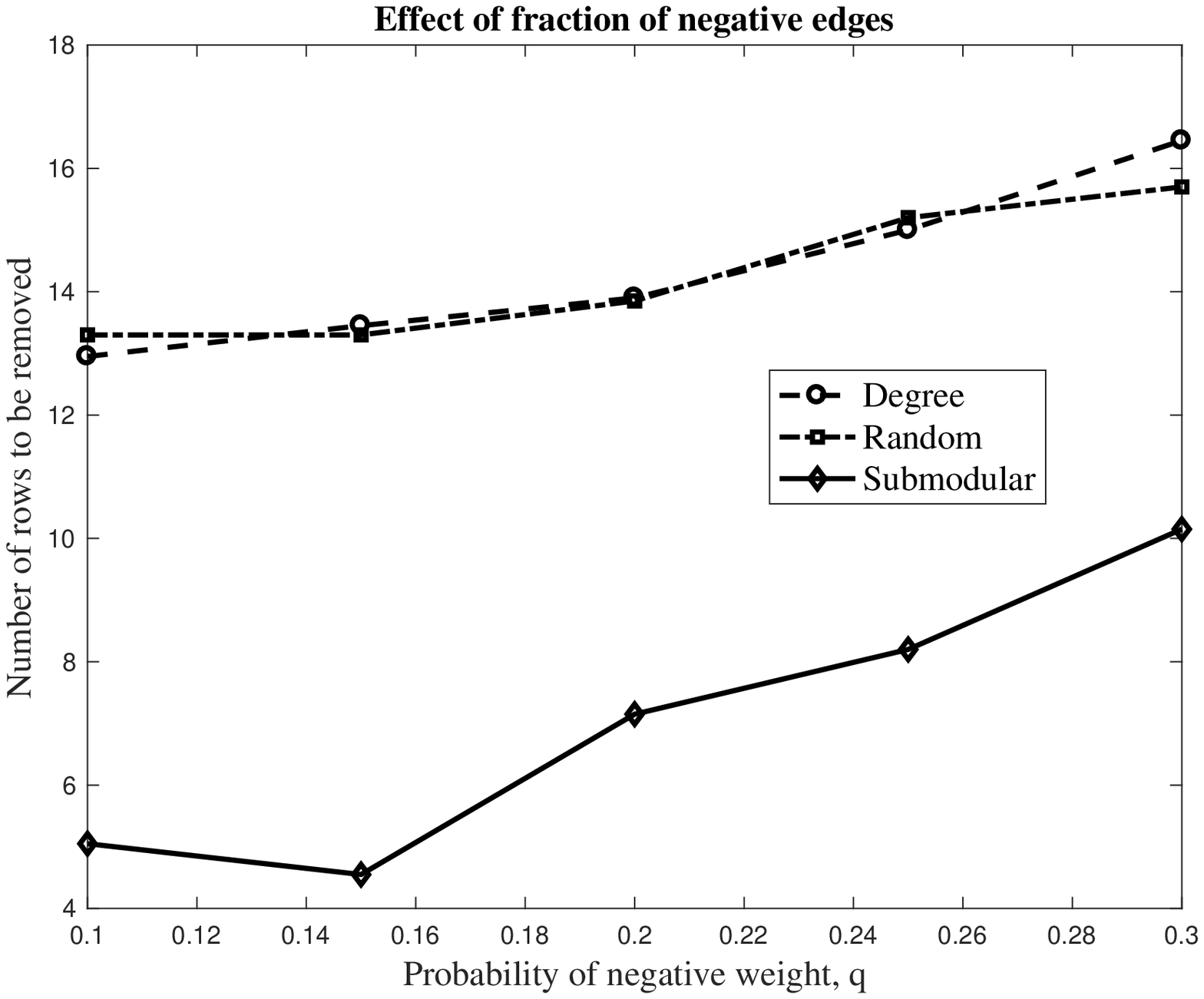} & 
\includegraphics[width=2in]{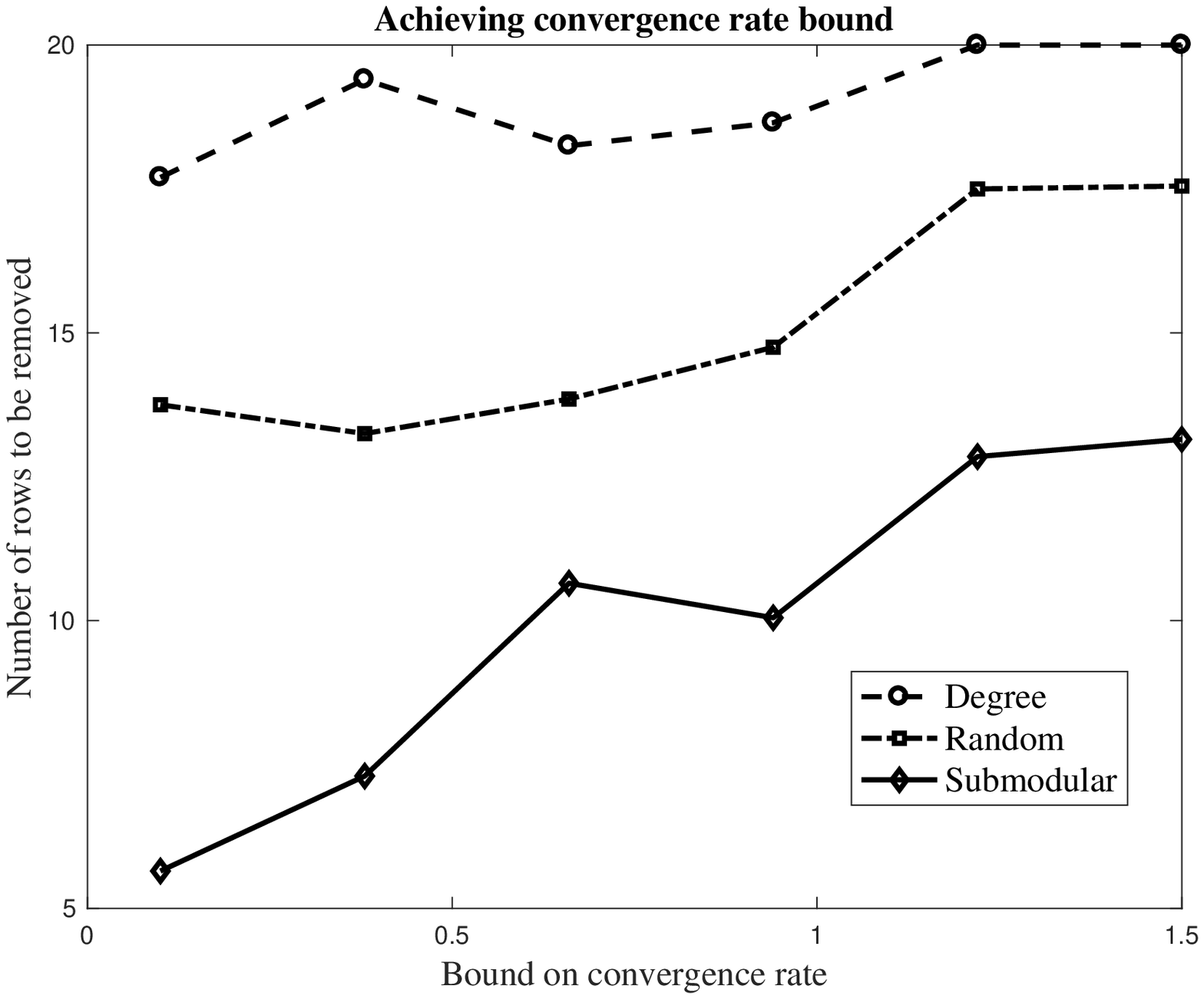}\\
\mbox{(a)} & \mbox{(b)} & \mbox{(c)}
\end{array}$
\caption{Numerical evaluation of Algorithm \ref{algo:greedy}. (a) Number of rows that must be removed as the network size increases in a geometric random graph with negative edges. The number of rows removed increases linearly for all methods, but the submodular optimization approach requires fewer rows to be removed. (b) Increasing the probability of a negative weight causes an increase in the number of rows to be removed for all methods, with the submodular approach consistently requiring fewer rows to be removed. (c) Number of rows that must be removed to achieve convergence rate bounds, as quantified by the smallest eigenvalue. Achieving faster convergence requires additional rows and columns to be removed.}
\label{fig:simulation}
\end{figure*}

The effect of the network size is shown in Figure \ref{fig:simulation}(a). As the number of network nodes increases, the number of rows that must be removed from the matrix to ensure consensus based on the submodular approach increases from 6 to roughly 25. This is significantly fewer than the number of rows that must be removed based on both the random and degree-based methods.


The effect of increasing the probability that an edge is negative is shown in Figure \ref{fig:simulation}(b) for a network of $20$ nodes. Adding negative edges reduces the eigenvalues of the Laplacian, and hence requires additional rows to be removed in order to provide stability. The submodular approach consistently requires fewer rows to be removed to compared to the other heuristic. We also observed that fewer random rows needed to be removed compared to removing rows with large degree.  

Figure \ref{fig:simulation}(c) shows the number of rows that must be removed in order to achieve given bounds on the minimum eigenvalue for networks that do not have negative edges in a network of $20$ nodes. More rows must be removed in order to provide a higher convergence rate. As in Figures \ref{fig:simulation}(a) and \ref{fig:simulation}(b), the submodular approach required fewer inputs than the random and degree-based methods. 
\section{Conclusions and Future Work}
\label{sec:conclusion}
This paper considered the problem of selecting a positive definite submatrix of a symmetric matrix. This problem arises naturally in contexts including maximizing convergence of distributed control algorithms and ensuring consensus in the presence of negative edges. We developed a submodular optimization approach to selecting a maximum-size submatrix. Our approach was based on proving that positive definiteness 
can be characterized through the probability distribution of the quadratic form induced by the matrix. We then proved that the derived condition is equivalent to a constraint on a submodular function, implying that satisfying a given bound on the eigenvalues is inherently a submodular optimization problem. We presented efficient greedy algorithms and analyzed the computational complexity. Extensions to non-symmetric matrices were discussed. We provided alternative sufficient conditions for Laplacian matrices based on the inverse trace and log determinant, and proved that these conditions can be expressed as submodular optimization problems by exploiting spectral submodular properties. Our approach was verified through numerical study.

The optimality bounds that we derived are parameter dependent, and are influenced by the matrix spectrum. Future work will attempt to characterize and eventually remove these dependencies. We also plan to investigate the problem of selecting a fixed-size submatrix in order to maximize the minimum eigenvalues. Finally, we will study generalizations to other application domains where the convergence rate of distributed algorithms, such as distributed convex optimization, is determined by graph eigenvalues.


\bibliographystyle{plain}        
\bibliography{autosam}           



\end{document}